\documentclass[12pt]{article}
\usepackage{amssymb}
\usepackage{amsmath}
\usepackage{cite}
\usepackage{epsf,graphicx}

\newcommand\ZZ{\hbox{\zfont Z\kern-.4emZ}}
\font\zfont = cmss10 %scaled \magstep1

\newcommand{\drawsquare}[2]{\hbox{%   
\rule{#2pt}{#1pt}\hskip-#2pt%  left vertical   
\rule{#1pt}{#2pt}\hskip-#1pt%  lower horizontal   
\rule[#1pt]{#1pt}{#2pt}}\rule[#1pt]{#2pt}{#2pt}\hskip-#2pt%  upper horizontal  
\rule{#2pt}{#1pt}}% right vertical   

\newcommand{\Yfund}{\raisebox{-.5pt}{\drawsquare{6.5}{0.4}}}%  fund   

\begin{document}
%%%%%%%%%%%%%%%%%%%%%%%%%%%%%%%%%%%%%%%%%%%%%%%%%%%%%%%%%%%%%%%%%%%%%%%%%%%%%%%%

\begin{titlepage}
%%%%%%%%%%%%%%%%%%%%%%%%%%%%%%%%%%%%%%%%%%%%%%%%%%%%%%%%%%%%%%%%%%%%%%%%%%%%%%%
\begin{flushright}
{\tt hep-th/0208095} \\
\end{flushright}

\vskip.5cm
\begin{center}
{\huge{\bf Seiberg-Witten Description of the Deconstructed 6D (0,2) 
Theory}}
\vskip.2cm
\end{center}
\vskip0.2cm

\begin{center}
{\sc Csaba Cs\'aki}$^{a}$,
{\sc Joshua Erlich}$^{b}$
{\sc and John Terning}$^{b}$ \\

\end{center}
\vskip 10pt

\begin{center}
$^{a}$ {\it Newman Laboratory of Elementary Particle Physics, Cornell University, Ithaca, 
NY 14853, USA } \\
\vspace*{0.1cm}
$^{b}$ {\it Theory Division T--8,
Los Alamos National Laboratory, Los Alamos, NM 87545, USA} \\
\vspace*{0.1cm}
{\tt  csaki@mail.lns.cornell.edu, erlich@lanl.gov, terning@lanl.gov}
\end{center}

\vglue 0.3truecm

\begin{abstract}
\vskip 3pt \noindent 
It has recently been suggested that, in a large $N$ limit, a
particular four dimensional gauge theory is indistinguishable from the
six dimensional CFT with (0,2) supersymmetry compactified on a torus.
We give further evidence for this correspondence by studying the
Seiberg-Witten curve for the ``deconstructed'' theory and
demonstrating that along the reduced Coulomb branch of moduli space (on the 
intersection of the Higgs and
Coulomb branches) it describes the low energy physics on a stack
of M5-branes on a torus, which is the (0,2) theory on a torus
as claimed.  The M-theory construction helps to clarify the enhancement
of supersymmetry in the deconstructed theory at low energies, and 
demonstrates its stability to radiative and instanton corrections. 
We demonstrate the role of the theta vacuum in the deconstructed theory.  
We point out that by varying the theta
parameters and gauge couplings in the deconstructed theory, 
the complex structure
of the torus can be chosen arbitrarily, and the torus is not
metrically $S^1\times S^1$ in general.

\end{abstract}

\end{titlepage}

\newpage

%\renewcommand{\thefootnote}{(\arabic{footnote})}

%%%%%%%%%%%%%%%%%%%%%%%%%%%%%%%%%%%%%%%%%%%%%%%%%%%%%%
%%%%%%%%%%%%%%%%%%%%%%%%%%%%%%%%%%%%%%%%%%%%%%%%%%%%%%
\section{Introduction}
\label{sec:intro}
\setcounter{equation}{0}
\setcounter{footnote}{0}
%%%%%%%%%%%%%%%%%%%%%%%%%%%%%%%%%%%%%%%%%%%%%%%%%%%%%%
%%%%%%%%%%%%%%%%%%%%%%%%%%%%%%%%%%%%%%%%%%%%%%%%%%%%%%

It has recently been demonstrated that certain four dimensional gauge theories
can reproduce the physics of higher dimensional gauge theories, typically up
to some energy scale  
\cite{deconstruction}.  The four dimensional theory is referred to as 
the ``deconstructed,'' or ``latticized,'' theory.  
Gauge theories in more than four dimensions are
typically not renormalizable, so the deconstructed theory provides an
ultraviolet completion for the higher dimensional theory.
The basic idea is to begin with a four dimensional 
action which, after certain elementary or composite
fields get vacuum expectation values (VEVs), is the action for the higher
dimensional theory latticized in all but four dimensions, plus corrections
which become small if the lattice spacing is relatively small.  The lattice
spacing can generally not be taken to zero in a controlled way
because this requires the infinite coupling limit of the gauge theory.  
Deconstruction of extra dimensions has proved useful in model 
building 
\cite{CEGK,0-2,BFLP,KKU,ACG,modelbuilding,GUT,wittengut,witek,littlehiggs}, 
and  often the number of lattice sites can be taken as
small as two while maintaining the relevant features of the extra 
dimensions.  

Deconstruction of supersymmetric (SUSY) models is especially
interesting \cite{CEGK,0-2,BFLP,KKU}. The deconstructed SUSY theory
typically has less supersymmetry at high energies than at the low
energies for which it describes the higher dimensional model.  This unusual
behavior is
due to an accidental symmetry on the chosen branch of moduli space of
the theory, and it must be checked that there are no radiative or
nonperturbative corrections which would violate the enhancement of
SUSY.  In the case where the 5D SUSY gauge theory with 8 supercharges
is deconstructed to a theory with ${\cal N}$=1 SUSY in 4D (4
supercharges) \cite{CEGK}, stability of the spectrum and other holomorphic data
follows from a
non-anomalous U(1)$_R$ symmetry which is preserved, or alternatively
by the ${\cal N}$=1 nonrenormalization theorem which prevents the
generation of ${\cal N}$=2
SUSY breaking terms in the superpotential.  Here we will
study a six dimensional theory with 16 supercharges, deconstructed to
a theory with 8 supercharges.  In this case we will explicitly
understand the stability of SUSY enhancement from the Seiberg-Witten
description of the theory.  This model was studied in \cite{0-2}, and
we will generalize the model to the case with arbitrary theta vacuum
in the deconstructed theory.  By varying the theta vacuum we
deconstruct the 6D theory with (0,2) SUSY compactified on a torus with
arbitrary complex structure.

The model is a 4D SU$(M)^N$ gauge theory with ${\cal N}$=2 SUSY and $N$
bifundamental hypermultiplets $Q_i$ as follows: \begin{equation}
\label{fieldcontent}
\begin{array}{c|cccccc}
 &SU(M)_1&SU(M)_2&SU(M)_3&\cdots&SU(M)_{N-1}&SU(M)_N \\ \hline
Q_1&\Yfund& \overline{\Yfund}&1&\cdots&\cdots&1 \\
Q_2&1&\Yfund& \overline{\Yfund}&\cdots&\cdots&1 \\
  &\vdots&\vdots&\vdots&\ddots&\ddots&\vdots   \\
Q_{N-1}&1&\cdots&\cdots&\cdots&\Yfund&\overline{\Yfund} \\ 
Q_N&\overline{\Yfund}&1&\cdots&\cdots&1&\Yfund
\end{array}
\end{equation}
$Q_i$ represents a vectorlike pair of ${\cal N}$=1 
chiral multiplets in (\ref{fieldcontent}).
The SU$(M)$ gauge couplings $g$ are all taken to be equal, and
we will usually assume that the theta parameters for
individual SU$(M)$ gauge group factors are equal.  This theory has a
Higgs branch, for which the scalars in the $Q_i$ hypermultiplets get
VEVs, and a Coulomb branch, for which the adjoint scalars in the
${\cal N}$=2 vector multiplets get VEVs and the gauge group is
generically broken to U$(1)^{N(M-1)}$.  If the $Q_i$ all have equal
VEVs $v{\mathbf 1}$ (and the adjoints do not have VEVs), then the
gauge group is broken to a diagonal SU$(M)_D$.  At energies much
smaller than $gv/N$ the theory is 4D ${\cal N}$=4 SU$(M)$ gauge
theory.  If the theta parameters vanish and $N\gg 1$ with $g^2/N$ and
$gv/N$ held fixed, then it was argued in \cite{0-2} that the low lying
dyonic spectrum is the KK spectrum of 6D (0,2)$_M$ superconformal
field theory on a torus of radii $N/(2\pi gv)$ and $g/(8\pi^2 v)$.
The correspondence was made precise by starting with a Type IIB string
theory description of the gauge theory and showing that the
corresponding M theory description on a baryonic Higgs branch ({\em
i.e.} when the ``baryons'' $Q_i^M$ have expectation values, 
in analogy with the baryonic branch in SUSY QCD \cite{APS}) is that of a
stack of $M$ M5-branes wrapped on the torus, with the 11D Planck length
$l_P$ taken to zero.  This is the (0,2)$_M$ theory  \cite{Schwarz}
compactified on a torus.  In the case $M\rightarrow\infty$ the
deconstructed theory is a holographic description of Type IIB string
theory on a pp-wave background \cite{MRV}.

We will study the correspondence between the (0,2) theory and its
deconstructed version by starting with the Type IIA description of
the deconstructed theory, and raising
it to M-theory before moving along the baryonic Higgs branch.   
This allows us to easily determine the complex and metric
structure of the torus in terms of the parameters of the deconstructed theory.
The M-theory description involves an M5-brane wrapped on the Seiberg-Witten
curve which describes the dynamics of the gauge theory on the Coulomb branch.
The Seiberg-Witten description of an ${\cal N}$=1 version of this
${\cal N}$=2 theory was studied in \cite{CEKPSS}.
The Seiberg-Witten curve in the ${\cal N}$=2 case studied here 
will be useful despite the fact that the relevant branch of moduli space for 
deconstruction is the Higgs branch, not the Coulomb branch. 

There is a baryonic reduced Coulomb branch in which both the $Q_i$ and
adjoints have VEVs.
The gauge group in the low energy theory 
along this branch is U(1)$^{M-1}$.
The Coulomb and reduced Coulomb branches meet when the
bifundamental VEV $v$ vanishes and the adjoint scalar VEVs among the
$N$ gauge group factors are equal.  
Along the reduced Coulomb branch, 
the Coulomb and Higgs sectors have a product structure.  In other words, 
$Q_i$ VEVs do not affect the low energy behavior of the adjoint fields, and
vice versa.
This follows by studying constraints on the K\"ahler potential 
from ${\cal N}$=2 supersymmetry \cite{APS}.  

The Seiberg-Witten curve becomes singular and factorizes on the intersection
of the Higgs and Coulomb branches \cite{GP,ENR}.  One factor
precisely describes the complex structure of the 
torus on which the M5-branes are wrapped, which
allows us to easily identify the torus in terms of the gauge theory 
parameters. By the decoupling of the Higgs and Coulomb branches, the complex
structure of the torus does not depend on the $Q_i$ VEVs.  This is true
even along the root of the reduced Coulomb branch,
when the adjoint VEVs vanish and the low energy theory has an
SU$(M)$ gauge symmetry.  This is the domain of moduli space 
which describes the
(0,2)$_M$ theory.  Hence, the Seiberg-Witten curve contains information
about the (0,2)$_M$ theory on the torus.  The low energy dynamics of
the (0,2) theory on a torus, which is 4D ${\cal N}$=4 SU$(M)$ gauge theory, 
is recovered exactly without corrections from the additional moduli involving
the $Q_i$ hypermultiplet VEVs.
The factorization of the Seiberg-Witten curve is easy to understand from the
M-theory 5-brane description of the theory.  It 
also demonstrates that the enhancement of SUSY at low energies is stable to
radiative and instanton corrections.  
A more detailed analysis is the purpose of this paper.

%%%%%%%%%%%%
\section{Type IIA and M-theory construction of 
the deconstructed (0,2) theory}
\label{sec:SW}
\begin{figure}[t]
\epsfxsize=1.5in
\centerline{\epsfbox{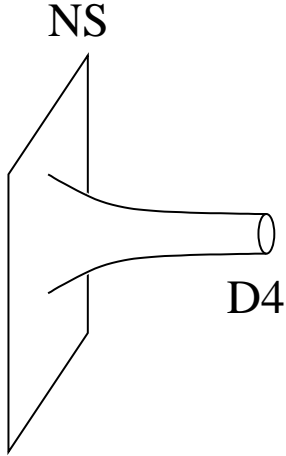}}
\caption{The intersection of a Type IIA D4-brane and NS5-brane is 
generically smoothed out into a single M5-brane in M-theory.}
\label{fig:smooth} \end{figure}

\setcounter{equation}{0}
\setcounter{footnote}{0}
%%%%%%%%%%%
Witten demonstrated that the low energy dynamics of a class of ${\cal
N}$=2 supersymmetric gauge theories, which arise as the low energy theories
on configurations
of NS5-branes and D4-branes of Type IIA string theory, can be understood
by lifting the corresponding brane configurations to M-theory
\cite{Witten}.  Intersections
of D-branes and NS5-branes are generally smoothed out into a single M5-brane,
as in Figure~\ref{fig:smooth}.
In this interpretation, the D4-brane corresponds to the M5-brane wrapped on
the M-theory circle of radius $R_{10}$, 
and the NS5-brane corresponds to the M5-brane at a point
on the M-theory circle.  The algebraic curve which describes the geometry of
the M5-brane was shown in \cite{Witten} to be precisely the Seiberg-Witten
curve \cite{SW}
from which the holomorphic gauge coupling $\tau=4\pi i/g^2 +\theta/2\pi$
and K\"ahler potential are 
determined along the Coulomb branch of the corresponding gauge theory.
%${\cal N}$=2 theory which lives
%on the brane configuration at low energies.

\begin{figure}[t]
\epsfxsize=4.5in
\centerline{\epsfbox{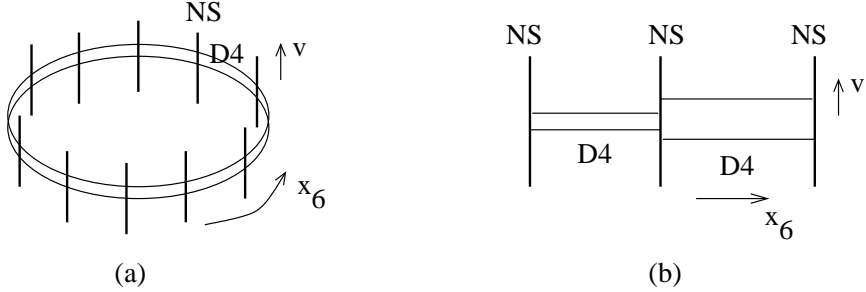}}
\caption{Type IIA brane configurations for theories studied here. (a)
Deconstructed
(0,2) theory. There are $N$ NS5-branes and $M$ D4-branes between
neighboring NS5 branes. (b) ${\cal N}$=2
SU$(M)\times$ SU$(M)$ gauge theory with bifundamental hypermultiplet. There
are three NS5 branes and $M$ D4-branes between neighboring NS5-branes.}
\label{fig:branes}\end{figure}

The 4D theory proposed in \cite{0-2} as a deconstructed version of the
six dimensional (0,2) theory is derived from a Type IIA brane
configuration as in Figure~\ref{fig:branes}a.  
There are $N$
NS5-branes which live in the $(x_0,\dots,x_3;x_4,x_5)$ directions and
$M$ D4-branes stretched along the $(x_0,\dots,x_3;x_6)$ directions
between neighboring pairs of NS5-branes.  The $x_6$ direction is
compactified on a circle of radius $R_6$, which will be identified by the
KK spectrum.  All of the branes are at
$x_7=x_8=x_9=0$ to begin with.  The Type IIA description is T-dual to the Type
IIB $A_{N-1}$ quiver theory described, for example, in \cite{DM}.
The gauge theory that lives on the brane configuration described above, at 
energies below the string scale $l_s^{-1}$, is
an ${\cal N}$=2 SU$(M)^N$ gauge theory with bifundamental hypermultiplets 
$Q_i$ as given in (\ref{fieldcontent}).
This is the theory described in the Introduction.  In fact, there is an
additional U(1) in the theory, but nothing is charged under it so it
decouples from the theory we are interested in. The gauge coupling of the 
$i^{th}$ group factor is given by,
\begin{equation}
\frac{4 \pi}{g_i^2}=\frac{x_6^i-x_6^{i-1}}{g_s l_s},
\end{equation}
where $g_s$ is the string coupling, and $l_s$ is the string length.
The branch of moduli
space relevant to the deconstructed (0,2) theory is along the Higgs
branch with all the $Q_i$ VEVs equal and proportional to the identity,
$Q_i=v\mathbf{1}$.  On this branch the gauge group is broken to a
diagonal SU$(M)_D$ with a massless adjoint hypermultiplet.  The
adjoint scalar in the SU$(M)_D$ vector multiplet can still get a VEV
in the Cartan subalgebra of SU$(M)_D$,
which would generically break the gauge group further to U(1)$^{M-1}$.
If the adjoint gets a generic
VEV this is a reduced Coulomb branch, an analog of the baryonic branch
in \cite{APS}. The positions of the
D4-branes in the $z=x_4+ix_5$-direction correspond to
the VEVs of the adjoint scalars in the SU$(M)^N$ vector multiplet (up
to instanton corrections).  If the adjoint VEVs are equal for all of
the SU$(M)$ gauge group factors, then the D4-branes wrap the circle
and decouple from the NS5-branes, so that the NS5-branes can be
removed while remaining on the vacuum manifold.  (This will be made explicit
in the next section.)  This is the
intersection of the Coulomb and reduced Coulomb branches.  The
bifundamental hypermultiplets correspond to strings which cross the
NS5-branes, and they obtain VEVs when the NS5-branes are removed from
the circle in the $x_7,\, x_8, \,x_9$ directions.  When all of the
NS5-branes are removed from the D4-brane circle, the gauge theory on
the circle is generically broken to U(1)$^{M-1}$.  This is the reduced Coulomb
branch discussed earlier. 

The relevant limit of the deconstructed theory is 
$l_s\to 0$ (in order to decouple the string modes so that the brane 
dynamics is described by the field theory ), $N\to \infty$,
$g\to \infty$, $v\to \infty$, $g_s= g/(8\pi^2 v l_s)\to \infty$ such that 
$N/gv={\rm const}$, $g_s l_s={\rm const}$. These limits imply that the theory is 
better described within M-theory compactified on a circle $R_{10}= g_s l_s,$ 
which is held fixed in the deconstruction limit.
When lifted to M-theory the brane
configuration is generically described by a single M5-brane wrapped on the
Seiberg-Witten curve.  The nontrivial geometry of the M5-brane is
confined to the directions $x_4,x_5,x_6$ and the M-theory direction
$x_{10}$.  The effective four-dimensional low energy theory lives in
the directions $x_0,\dots,x_3$.
The D4-branes become M5-branes which are
wrapped around both the M-theory circle in the $x_{10}$ direction and
the circle in the $x_6$-direction.  For small $l_s$ and constant $R_{10}$, 
the 11D Planck length $l_P^3=l_s^2 R_{10}$ is small, and
the theory which lives on this
brane configuration is the 6D (0,2) theory wrapped on a
torus up to energies which probe the NS5-branes.  We will discuss the
relevant energy scales in more detail later.
At energies smaller than the inverse size of the torus, the theory
is simply 4D ${\cal N}$=4 Yang-Mills theory.  
This demonstration was another approach to the
same M5-brane configuration described in \cite{0-2}.  But, as we will see
by studying the Seiberg-Witten curve for this theory, the choice of
theta vacuum of the deconstructed theory is reflected in the complex
structure of the manifold on which the M5-brane is wrapped.
%, which is in general metrically not a product $S^1\times S^1$.

%%%%%%%%%%%%
\section{Factorization of the Seiberg-Witten curve and enhancement of supersymmetry}
\label{sec:factorization}
\setcounter{equation}{0}
\setcounter{footnote}{0}
%%%%%%%%%%%

In this section we will demonstrate the factorization of the Seiberg-Witten curve
on the reduced Coulomb branch, and thus the enhancement of supersymmetry.
To begin we will avoid the additional complications which arise from
putting the $x_6$ direction on a circle, and first demonstrate in more
detail the decoupling of the Coulomb and Higgs branches in the
Seiberg-Witten curve for the following ${\cal N}$=2 theory with hypermultiplet 
$Q$:
\[
\begin{array}{c|cc}
&SU(M)_1&SU(M)_2 \\ 
\hline Q&\Yfund& \overline{\Yfund}
\end{array} \]
Recall that the hypermultiplet is vectorlike, so this theory is anomaly
free.  It is described by the brane configuration shown in
Figure~\ref{fig:branes}b.  Note that this theory differs from that
studied in \cite{IS,CEFS} in that the latter had a vanishing
superpotential, whereas the ${\cal N}$=2 theory studied here has
superpotential, \begin{equation}
W=Q_A^{\ \alpha}\,\phi_\alpha^{\ \beta}\,\widetilde Q_\beta^{\ A}-\widetilde
Q_\alpha ^{\ A}\,\varphi_A^{\ B}\,Q_B^{\ \alpha}+\lambda\,
\phi_\alpha^{\ \alpha}+ \mu\,\varphi_A^{\ A}, \label{eq:W}
\end{equation}
where $Q$ and $\widetilde Q$ are the two ${\cal N}$=1
chiral multiplets in the bifundamental
hypermultiplet, and $\phi$ and $\varphi$ are the adjoint chiral multiplets 
of the two
SU$(M)$ gauge groups.  Greek and Roman indices run from $1$ to $M$ and
refer to the two gauge group factors. 
The Lagrange multipliers $\lambda$ and $\mu$ enforce
the tracelessness of the adjoint fields.  Although the superpotential
was absent in \cite{CEFS}, their analysis of the $D$-flatness constraints
on $Q$ and $\widetilde Q$ is unchanged as long as the adjoints $\phi$ and
$\varphi$ do not get VEVs.  
It was found there that the general solution to the $D$-flatness
constraints was that $Q$ and $\widetilde Q$ take the form, \begin{equation}
Q=\left(\begin{array}{ccc}
q_1&& \\
&\ddots& \\
&&q_N \end{array}\right),\ \ \ \ \widetilde Q=\left(\begin{array}{ccc}
\sqrt{q_1^2+c}&& \\
&\ddots& \\
&&\sqrt{q_N^2+c} \end{array}\right)\,C\,e^{i\beta},\end{equation}
where $C$ is a diagonal SU$(M)$ matrix.\footnote{A complete analysis of
the moduli space of a more general class of theories which includes this one
can be found in \cite{GP}.}

The $F$-flatness constraints can be read off of the superpotential
(\ref{eq:W}), namely
\begin{eqnarray}
\lambda\,\delta_\beta^{\ \alpha}+\widetilde Q_\beta^{\ A}\,Q_A^{\ \alpha}  = 0 \\
Q_A^{\ \alpha}\,\phi_\alpha^{\ \beta}-\varphi_A^{\ B}\,Q_B^{\ \beta} =0
\end{eqnarray}
and two similar equations of motion with respect to
$\varphi$ and $Q$.  Note that generically the bifundamentals $Q$ 
and $\widetilde Q$
cannot get VEVs if the adjoints $\phi$ and
$\varphi$ do have VEVs, but if $\phi$
and $\varphi$ have equal nonvanishing VEVs in the Cartan subalgebra
of the corresponding SU$(M)$, then $Q$ and $\widetilde Q$ can get VEVs
proportional to the identity.  In particular, we can choose 
$Q=\widetilde Q=v\mathbf{1}$.  This is the reduced Coulomb branch
discussed earlier.

We can easily recover this behavior from the Seiberg-Witten curve for this
theory.  The Seiberg-Witten curve and
one-form determine the holomorphic gauge coupling function and
K\"ahler potential on the Coulomb branch as a function of the Coulomb branch
moduli of the  theory, namely the
VEVs of gauge invariant operators \cite{LT}, \begin{eqnarray}
u_n^{(1)}&=&\left<{\rm Tr}\, \phi^n\right> \label{eq:moduli1} \\   
u_n^{(2)}&=&\left<{\rm Tr}\, \varphi^n\right>\,, \label{eq:moduli2}
\end{eqnarray} 
where $n=2,\dots,M$.  But we
will restrict the VEVs to all be equal so that we intersect the Higgs
branch, in which case there are only $M-1$ 
independent moduli in the curve, as opposed to $N(M-1)$.\footnote{On the
reduced Coulomb branch there are additional moduli corresponding to the 
hypermultiplet VEVs.  However, they cannot appear in the Seiberg-Witten
curve because of the product structure of the Higgs and Coulomb branches
\cite{APS}.}  
If we define the
complex parameters $z=x_4+ix_5$ and $t=\exp(x_6+ix_{10}/R_{10})$ 
(which is holomorphic because $x_{10}\sim x_{10}+ 2\pi R_{10}$), then Witten
argued that the M5-brane is described by the curve \cite{Witten},
\begin{equation}
\Lambda_1^M t^{3}+ A_1(z)t^2+\cdots+A_2(z)t + \Lambda_2^M =0,
\label{eq:SW2}\end{equation}
where $A_i(z)$ is a polynomial of degree $M$ in $z$, and $\Lambda_i$ is the
dynamical scale of SU$(M)_i$.  At constant $t$, 
(\ref{eq:SW2}) has $M$ roots for $z$, which correspond to the positions
of the $M$ D4-branes at that $t$.  At constant $z$, (\ref{eq:SW2}) has
$3$ roots for $t$, which correspond to the positions of the NS5-branes
at that $z$.

If we write, \begin{equation}
A_i(z)=\prod_{j=1}^M (z-a_{i,j}), \end{equation}
then the $a_{i,j}$ ($i=1,2$; $j=1,\dots,M$) represent the positions of the 
$M$ D4-branes between the
$i$th and $(i+1)$th NS5-branes. 
In terms of the SU$(M)\times$SU$(M)$ 
gauge theory
that lives on the branes, for $i=1,2$, the $a_{i,j}$ are the 
VEVs of the adjoint in the $i$th SU$(M)$ factor 
(up to instanton corrections)\cite{Witten}.  
The difference in the
average of the $a_{i,j}$ on the two sides of an NS5-brane is the bare
mass of the corresponding bifundamental hypermultiplet, which we take
to be zero, \begin{equation}
\frac{1}{M}\sum_{j=1}^{M}\left(a_{1,j}-a_{2,j}\right) = 0.
\end{equation}
The adjoint VEVs are in the Cartan subalgebra of the gauge group, but
are not gauge invariant.  They transform under the Weyl group
(which preserves the Cartan subalgebra),
which permutes the VEVs $a_{i,j}$ for fixed $i$.  Instead, the curve should 
be written in terms of the gauge invariant moduli, which are in general
shifted from their classical values by instanton effects.  
The Seiberg-Witten curve can then be written
\cite{ENR},
\begin{equation}
\Lambda_1^M\,t^3+\left(P_1(z;u_n^{(1)},\Lambda_1^M)+\Lambda_1^M\right)t^2+
\left(P_2(z;u_n^{(2)},\Lambda_2^M)+\Lambda_2^M
\right) t+\Lambda_2^M, \label{eq:SW2a}\end{equation}
where $P_1(z)$ and $P_2(z)$ are the same function of $z$, the 
Coulomb branch moduli $u_n^{(i)}$ defined in (\ref{eq:moduli1}) and
(\ref{eq:moduli2}),
and the dynamical scale of the corresponding SU$(M)_i$ group 
factor.\footnote{$P_i(z,u_n^{(i)},\Lambda_i^M)$ 
contains a constant term proportional to $\Lambda_i^M$.
This represents a shift in the maximal moduli $\left<{\rm Tr}\,\phi^M\right>$
and  $\left<{\rm Tr}\,\varphi^M\right>$ from their classical values.  We have
ignored a numerical constant in the definition of the dynamical scale.
For more on these subtleties, see for example \cite{ENR,GP}.}
Equating the VEVs of
the two adjoint scalars amounts to setting $P_1(z;u_n^{(1)},\Lambda_1^M)=
P_2(z;u_n^{(2)},\Lambda_2^M)$.  In that case,
the curve (\ref{eq:SW2a}) factorizes as, \begin{equation}
\Lambda_1^M(t+1)\left(t^2+\frac{P(z)}{\Lambda_1^M}t+\frac{\Lambda_2^M}{
\Lambda_1^M}\right) = 0. \end{equation}
If all moduli and dynamical scales are the same
for each gauge group factor SU$(M)_i$, this amounts to enforcing a reflection
symmetry about the middle NS5-brane.  If we fix $x_{10}=0$, the 
symmetry $x_6\rightarrow -x_6$ becomes
$t\rightarrow 1/t$.  If we then equate the dynamical scales $\Lambda_1$ and
$\Lambda_2$, and fix the moduli of both gauge groups to be equal, then 
the factorization of the curve immediately follows from the
$t\rightarrow 1/t$ symmetry.
Physically, the factorization corresponds 
to the fact that the hypermultiplet $Q$ can get a 
VEV from such a point in moduli space,
where the Coulomb and Higgs branches intersect, and the Higgs and Coulomb
branches decouple \cite{ENR}.
In the M-theory picture giving a VEV to $Q$ corresponds to removing the
middle brane, which is at $t=-1$ as we have written the curve, in the
$x_7,\,x_8,\,x_9$ directions.
By rescaling the curve by $t\rightarrow \Lambda_1^2\,t$, the dynamics along 
the 
reduced Coulomb branch is given by the factor multiplying $(t+1)$, which
we recognize to be simply the Seiberg-Witten curve for the diagonal 
SU$(M)_D$ gauge theory, with
dynamical scale $\Lambda_D^{2M}=\Lambda_1^M\Lambda_2^M$.  
That is, the Seiberg-Witten
curve describes the unbroken
U(1)$^{M-1}$ gauge theory on the reduced Coulomb branch.  Giving a VEV to the
hypermultiplet $Q$ corresponds to the removal of the middle brane in the
$x_7,\,x_8,\,x_9$ directions, but as mentioned earlier that dynamics is 
decoupled from the dynamics described by the Seiberg-Witten curve at low 
energies.

We now see explicitly from the brane construction that the enhancement
of SUSY along the reduced Coulomb or Higgs branch is stable to radiative and
nonperturbative corrections.  The D4-branes
break half of the 32 supersymmetries of the Type IIA string theory, and the
NS5-branes break another half.
If the NS5-branes are removed, 
then the remaining theory at energies below those which probe the NS5-branes
is that of the D4-branes on the
circle, which has twice as much SUSY.  In essence, the theory has factorized
into the theory on the parallel NS5-branes with 16 supercharges, and the
theory on the D4-branes with a different 16 supercharges.
The decoupling is confirmed by the factorization of the Seiberg-Witten curve.

As mentioned in the Introduction, we need to take the number of gauge group 
factors large in order to be able to probe the size of the torus because
otherwise the spectrum differs from the expected KK spectrum at low energies,
and the deconstructed theory can not correspond to a six dimensional theory
on a torus.  We can understand this from the string theory construction
as well by considering the relevant
energy scales.  The Type IIA brane configuration is useful in
understanding the gauge theory on that brane configuration at energies
less than $l_s^{-1}$.  The (0,2) theory is the theory on a stack of
M5-branes with vanishing 11D Planck scale, $l_P^3=l_s^2\,R_{10}$,
so we take $l_s\rightarrow 0$.  Furthermore, the theory that lives on the
wrapped M5-branes is only the
(0,2) theory up to scales where they probe the M5-branes that were removed
from the torus.
At energies much smaller than the inverse sizes of the
torus, the (0,2) theory becomes the 4D ${\cal N}$=4 theory.  We want to 
arrange the parameters of the theory so 
that we can probe somewhat higher energies, on the scale of the inverse
sizes of the torus, without also probing the NS5-branes.
Strings which stretch from a D4 brane to an NS5 brane and back carry
energy of order $gv$, corresponding to the fact that at that energy the
gauge theory
probes the individual gauge group factors and hypermultiplets $Q_i$.  

The size of the torus is determined
by the low lying ``electric''
spectrum of the gauge theory, which corresponds to the
KK modes around the $x_6$ circle.  This fixes $R_6=N/(2\pi gv)$, so we
must have $N$ large in order for the scale of low lying KK modes to be
much smaller than the bifundamental hypermultiplet scale $gv$. 
Notice
that the size of the torus is related to the positions of the NS5-branes. 
%in order to match to the field theory.
$R_{10}$
will be related to $R_6$ by identifying the low energy gauge coupling in
terms of $R_6/R_{10}$ via the M5-brane brane construction.
Alternatively, we can identify $R_{10}$ via the magnetic spectrum, as in
\cite{0-2}, which is interpreted as the KK modes around the other cycle
of the torus.  For nonvanishing theta parameter in the low energy SU$(M)_D$
gauge group, the low lying states in the large $N$ limit have masses,
\begin{eqnarray}
M_{n,m}=\frac{gv}{N}&\left|n+m\left(\frac{\theta}{2\pi}+i\frac{4\pi}{g_D^2}
\right)\right|& \nonumber \\
=\frac{1}{2\pi R_6}&\left|n+m\left(\frac{\theta}{2\pi}+i\frac{R_6}{R_{10}}
\right)\right|&, \label{eq:BPS}
\end{eqnarray} where $n$ and $m$ are the electric and magnetic 
quantum numbers of the BPS (in the sense of the theory with sixteen
supercharges) dyon, 
respectively, and correspond to the KK
numbers around the two one-cycles of the torus.
%and marginally
%bound states of such modes.  
%In the large $N$ limit, 
%there is an almost BPS charge
%lattice of states generated by the S-duality
%group for each electric KK level $k$, leading to a degeneracy of 
%states.\footnote{We are grateful to Andy Cohen for discussions on this point.}
Equation (\ref{eq:BPS}) also determines the complex structure of the
torus, $\tau=\theta/(2\pi)+i\,R_6/R_{10}$, in terms of the gauge theory
parameters.  Note that the bifundamental VEV $v$ sets the overall scale
for the flat metric on the torus, but does not appear in the period $\tau$.
This is once again a reflection of the product structure of the Higgs and
Coulomb branches where they intersect.
The first equality in (\ref{eq:BPS}) is the holomorphic extension of the
spectrum found in \cite{0-2}, where $v_h=gv$ is the holomorphic VEV.
The second equality in (\ref{eq:BPS}) follows immediately from
the M-theory construction as we will now see.

%%%%%%%%%%%%
\section{Compactification on an arbitrary torus: complex structure and 
factorization}
\label{sec:arbitrary}
\setcounter{equation}{0}
\setcounter{footnote}{0}
%%%%%%%%%%%

To identify the complex structure of the torus we will closely follow Witten
\cite{Witten}.  The M-theory circle in the $x_{10}$
direction has radius $R_{10}$ and the circle in the $x_6$ direction has radius
$R_6$.
The bare gauge coupling of the $i$th gauge group SU$(M)_i$ is given by the
difference in the $x_{6}$ positions of the $i$th and $(i+1)$th NS5-branes
\cite{HW,Witten},
\begin{equation}
\frac{4\pi}{g_i^2}=\frac{(x_{i+1}^{6}-x_i^{6})}{R_{10}}. \label{eq:g}
\end{equation}
The bare coupling of the diagonal gauge group is given by, \begin{equation}
\frac{4\pi}{g_D^2}=\sum_i \frac{4\pi}{g_i^2} = \frac{R_6}{R_{10}}. \end{equation}
The theta parameter of the SU$(M)_i$ gauge group factor is given by the 
difference in the $x_{10}$ positions of the $i$th and $(i+1)$th NS5-branes
\cite{Witten},
\begin{equation}
\theta_i=\frac{x_{i+1}^{10}-x_i^{10}}{R_{10}}. \label{eq:theta}\end{equation}
By ``going around the circle'' in the $x_6$ direction, namely by summing
(\ref{eq:theta}) over $\theta_i$, we would find\begin{eqnarray}
x_6&\rightarrow& x_6+2\pi R_6 \nonumber \\
x_{10}&\rightarrow& x_{10}+\theta R_{10}. \end{eqnarray}
By varying $\theta$, the theta parameters in the gauge theory can be chosen 
arbitrarily, as now $\sum_i \theta_i=\theta$.  The period of
the torus $\tau=\theta/2\pi + i R_6/R_{10}$ 
is identified with the 
%bare coupling and theta parameter 
holomorphic gauge coupling of the diagonal
SU$(M)_D$ gauge theory on the Higgs branch, $\tau=4\pi i/g_D^2+ \theta/2\pi$.
Thus, by varying $R_6/R_{10}$ and $\theta$, or equivalently by varying
the gauge couplings and theta parameters of the deconstructed
theory, the
complex structure of the torus on which the (0,2) theory is compactified
can be chosen arbitrarily.

Let us now discuss the factorization of the Seiberg-Witten curve  
on the intersection of the Higgs and
Coulomb branches for the full deconstructed theory. A classical analysis
similar to that done in Section~\ref{sec:factorization} indicates that
there is a reduced Coulomb branch in this theory along which the Seiberg-Witten
curve should factorize.
We 
write the Seiberg-Witten curve as in \cite{Witten}:
\begin{equation}
z^M-f_1(x,y)z^{M-1}+f_2(x,y)z^{M-2}+\cdots+(-1)^M\,f_M(x,y) = 0, 
\label{eq:SWcircle}
\end{equation}
where $x$ and $y$ parametrize the torus on which the M5-brane lives,
which is specified by an elliptic curve of the form, \begin{equation}
y^2=x^3+c_1 x+c_2, \label{eq:torus}\end{equation} for some complex
constants $c_1$ and $c_2$.  These constants, together with the unique 
nonsingular holomorphic differential $\omega=dx/y$ and a choice of cycles, 
determine the complex structure of the
torus.  To be precise, there is a basis of cycles $\gamma_1$ and
$\gamma_2$ on the torus with unit intersection from which the period of the
torus can be calculated as \begin{equation} \tau= \oint_{\gamma_1}
\omega \,\left(\oint_{\gamma_2} \omega\right)^{-1}.
\end{equation}
The period $\tau$ is to be identified with the gauge coupling and
$\theta$ parameter of the diagonal SU$(M)_D$ gauge group
via $\tau=4\pi i/g^2 + \theta/2\pi$.  The constants
$c_i$ in (\ref{eq:torus}) and the cycles $\gamma_i$ are chosen to give the
correct period $\tau$ on the moduli space of vacua.
There is also a flat metric on the torus which specifies its size.

For generic adjoint VEVs, there are $M$ roots for $z$, which describe the
positions of the D4-branes as before.  The functions $f_i(x,y)$ are
generically meromorphic functions on the torus which have simple poles.
At the pole  there would be a singularity in at least one of the roots of
$z$ which looks near the singularity like, \begin{equation}
(z-z_0)(s-s_0)=\epsilon, \label{eq:intersect} \end{equation}
where $s$ is a good local coordinate on the torus near the singularity.
This would describe the smooth connection of the D4-brane at $z_0$
to the NS5-brane at $s_0$.  But at points in the moduli space where the
Higgs and Coulomb branches meet, the curve factorizes.  This happens when
the $f_i(x,y)$ are constants.  The interpretation of this is as follows.
If we assume that $R_6\gg R_{10}$ then in the neighborhood of an NS5-brane
the Seiberg-Witten
curve is well described by the coordinates $z$ and 
$s={\rm exp}(x_6+i\,x_{10})$
with the $x_6$ position of the NS5-brane set to 0, say.
(Alternatively we can just focus on an appropriate neighborhood
of the NS5 brane for which $|x_6|\ll 2\pi R_6$,
and remove the restriction on $R_6/R_{10}$.)
When the Higgs and Coulomb branches intersect there is a symmetry 
corresponding to reflections about the NS5-brane, which as we saw earlier 
implies that the curve should factorize near the NS5-branes.  This implies
that locally the
$f_i(x,y)$ in (\ref{eq:SWcircle}) are constants, which by holomorphy implies
they are constants over the torus.  Then, using the S-duality which takes
$\tau\rightarrow -1/\tau$ and $\tau\rightarrow \tau+1$
we can relax the restriction $R_6\gg R_{10}$, so that the $f_i(x,y)$ are
generically constants on the intersection of the Higgs and Coulomb branches.

In order to completely specify the brane configuration in the case in which
it factorizes, the curve (\ref{eq:SWcircle}) should instead be written,
\begin{equation}
h_0(x,y)z^M-h_1(x,y)z^{M-1}+h_2(x,y)z^{M-2}+\cdots+(-1)^M\,h_M(x,y) =0,
\end{equation}
where now the $h_i(x,y)$ are all equal up to overall numerical constants.
All of the $f_i(x,y) = h_i(x,y)/h_0(x,y)$ 
of (\ref{eq:SWcircle}) are then constants.  The curve
factorizes as, \begin{equation}
h_0(x,y)\,\prod_{i=1}^M (z-z_i) =0, \end{equation}
where $z_i$ are the positions of the D4-branes, and the zeroes of 
$h_0(x,y)$ specify the positions of the NS5-branes.  The
singular intersections of
the D4-branes and NS5-branes are similar to (\ref{eq:intersect}) with
$\epsilon\rightarrow 0$.

Having understood how the M-theory 5-brane configuration contains information
about the reduced Coulomb branch that we are interested in, we can now
easily understand the emergence of the (0,2) theory from the deconstructed
theory.  
%The M5-brane wraps two circles, the M-theory circle and the
%circle in the $x_6$ direction.  
When the VEVs of all the adjoints are
equal, the Seiberg-Witten curve describing the brane configuration
factorizes into a product of terms reflecting the positions of the
flat NS5-branes and the positions of the
D4-branes which wrap the $x_6$ direction.
When the NS5-branes are removed symmetrically around the torus, 
the Seiberg-Witten curve describes
the remaining D4-branes, which are really M5-branes wrapped on the
$x_6$--$x_{10}$ torus.  If we also take the 11D Planck scale 
$l_P^3=l_s^2R_{10}$ to zero, then up to the presence of the additional 
NS5-branes this 
is precisely the (0,2)$_M$ theory on a torus whose complex structure we have
identified above.
At very low energies the theory on the D4-branes
is the 4D ${\cal N}$=4 SU$(M)$ gauge theory, the dynamics of which is not 
corrected by the presence of the NS5-branes (as a result of the product 
structure
of the Higgs and Coulomb branches).  At higher energies but below the
scale $gv$ which probes the positions of the NS5-branes, the spectrum
mimics the KK spectrum expected of the (0,2)$_M$ theory on a torus with the
same complex structure as that determined by the Seiberg-Witten curve.
This provides additional evidence that the deconstructed theory does 
indeed mimic the (0,2) theory
at scales which probe the size of the torus on which the (0,2) theory is
compactified.

%%***************************
\section*{Acknowledgments}
We are grateful to Allan Adams, Nima Arkani-Hamed, Andy Cohen, Mike Graesser,
Christophe Grojean, David B. Kaplan, Asad Naqvi,  
Luigi Pilo, Stephan Pokorski, Yuri Shirman, Matt Strassler and Arkady
Vainshtein
for useful conversations.  We are also happy to thank the Aspen Center for 
Physics, where this work was completed.
The research of C.C. is supported in part by the NSF, and in part by the
DOE OJI grant DE-FG02-01ER41206.  The research of J.E. and J.T. is supported
by the US Department of Energy under contract W-7405-ENG-36.

%%%%%%%%%%%%%%%%%%%%%%%%%%%%%%%%%%%%%%%%%%
\vspace{\baselineskip}

\end{document}